\documentclass[reprint,superscriptaddress,aps,prb,twocolumn,floatfix]{revtex4-2}
\usepackage[utf8]{inputenc}
\usepackage{textcomp} 
\usepackage{setspace}
\usepackage{amsmath}
\usepackage{breqn}
\usepackage{graphicx}
\usepackage{verbatim}

\usepackage{amsfonts}
\usepackage{amssymb}
\usepackage{xcolor}
\usepackage{soul}
\usepackage{tabularx}
\setcitestyle{super}
\usepackage{float}

\usepackage[normalem]{ulem}

\usepackage[colorlinks,linkcolor=blue,anchorcolor=blue,citecolor=blue,urlcolor=black]{hyperref}
\DeclareGraphicsExtensions{.pdf,.eps,.png,.jpg,.mps}

\begin{document}

\title{Probing material absorption and optical nonlinearity of integrated photonic materials}

\author{Maodong Gao$^{1,\ast}$, Qi-Fan Yang$^{1,\ast}$, Qing-Xin Ji$^{1,\ast}$, Heming Wang$^{1}$, Lue Wu$^{1}$, Boqiang Shen$^{1}$, Junqiu Liu$^{2}$, Guanhao Huang$^{2}$, Lin Chang$^{3}$, Weiqiang Xie$^{3}$, Su-Peng Yu$^{4}$, Scott B. Papp$^{4,\dagger}$, John E. Bowers$^{3,\dagger}$, Tobias J. Kippenberg$^{2,\dagger}$ and Kerry J. Vahala$^{1,\dagger}$\\
$^1$T. J. Watson Laboratory of Applied Physics, California Institute of Technology, Pasadena, California 91125, USA\\
$^2$Institute of Physics, Swiss Federal Institute of Technology Lausanne (EPFL), CH-1015 Lausanne, Switzerland\\
$^3$ECE Department, University of California Santa Barbara, Santa Barbara, CA 93106, USA\\
$^4$National Institute of Standards and Technology, Boulder, CO 80305, USA\\
$^{\ast}$These authors contributed equally to this work.\\
$^{\dagger}$Corresponding authors: scott.papp@nist.gov, jbowers@ucsb.edu, tobias.kippenberg@epfl.ch, vahala@caltech.edu}

\maketitle

{\bf\noindent Optical microresonators with high quality ($Q$) factors are essential to a wide range of integrated photonic devices.
Steady efforts have been directed towards increasing microresonator $Q$ factors across a variety of platforms. With success in reducing microfabrication process-related optical loss as a limitation of $Q$, the ultimate attainable $Q$, as determined solely by the constituent microresonator material absorption, has come into focus. Here, we report measurements of the material-limited $Q$ factors in several photonic material platforms. High-$Q$ microresonators are fabricated from thin films of SiO$_2$, Si$_3$N$_4$, Al$_{0.2}$Ga$_{0.8}$As and Ta$_2$O$_5$. By using cavity-enhanced photothermal spectroscopy, the material-limited $Q$ is determined. The method simultaneously measures the Kerr nonlinearity in each material and reveals how material nonlinearity and ultimate $Q$ vary in a complementary fashion across photonic materials. Besides guiding microresonator design and material development in four material platforms, the results help establish performance limits in future photonic integrated systems.}

Performance characteristics of microresonator-based devices improve dramatically with increasing $Q$ factor \cite{vahala2003optical}. Nonlinear optical oscillators, for example, have turn-on threshold powers that scale inverse quadratically with $Q$ factor \cite{spillane2002,kippenberg2004kerr,lee2012chemically}. The fundamental linewidth of these and conventional lasers also vary in this way \cite{STLinewidth,Vahala2008,Li2012}. In other areas including cavity quantum electrodynamics \cite{aoki2006observation}, integrated quantum optics \cite{lu2019chip,lu2019efficient,lukin20204h,ma2020ultrabright}, cavity optomechanics \cite{kippenberg2008cavity} and sensing \cite{vollmer2012review}, a higher $Q$ factor provides at least a linear performance boost. In recent years, applications that rely upon these microresonator-based phenomena, including microwave generation \cite{li2013microwave}, frequency microcomb systems \cite{Kippenberg2018}, high-coherence lasers \cite{Li2012,gundavarapu2019sub,jin2021hertz} and chip-based optical gyroscopes \cite{Cascaded_SBL_Gyro,Crystalline_Passive_Gyro,lai2020earth}, have accelerated the development of high-$Q$ photonic-chip systems \cite{ji2017ultra,zhang2017monolithic,yang2018bridging,wilson2020integrated,chang2020ultra,xie_ultrahigh-q_2020,liu2020photonic,jin2021hertz,liu_aluminum_2021,gao_broadband_2021, biberman_ultralow-loss_2012}. 

\begin{figure}[t!]
\centering
\includegraphics[width=\linewidth]{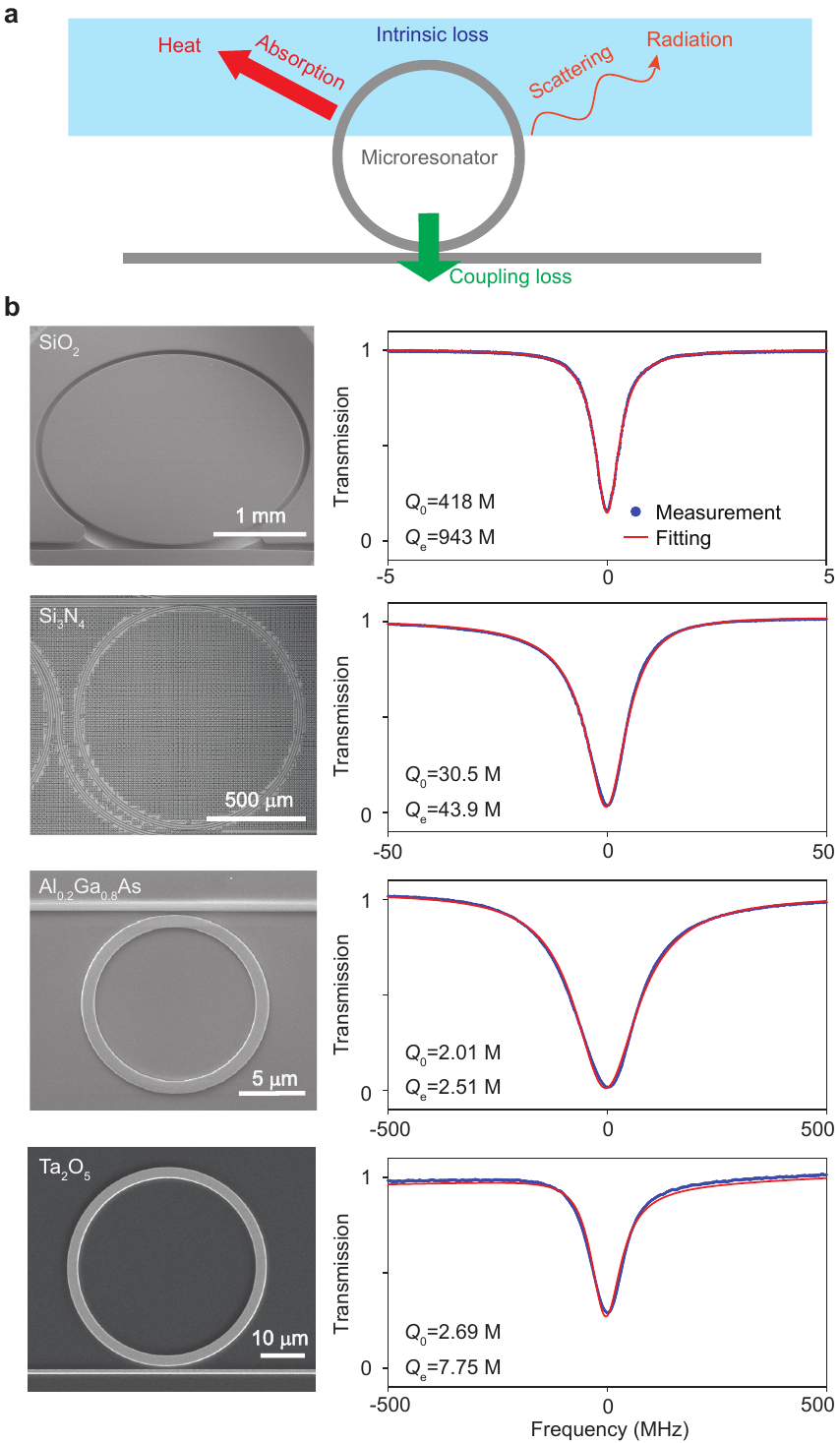}
\caption{{\bf 
High-$Q$ optical microresonators characterized in this work.}
{\bf a,} Schematic showing optical loss channels for high-$Q$ integrated optical microresonators. The loss channels include surface (and bulk) scattering loss and material absorption loss. The intrinsic loss rate is characterized by the intrinsic $Q$ factor ($Q_0$). Bus waveguide coupling also introduces loss that is characterized by the external (coupling) $Q$ factor ($Q_e$).
{\bf b,} Left column: images of typical microresonators used in this study. Right column: corresponding low input-power spectral scans (blue points) with fitting (red). The intrinsic and external $Q$ factors are indicated. M: million.
}
\label{figure1}
\end{figure}

$Q$ factor is determined by material losses, cavity loading (i.e., external waveguide coupling), and scattering losses (see Fig. \ref{figure1}a). To increase $Q$ factor, there have been considerable efforts focused on new microfabrication methods and design techniques that reduce scattering loss associated with interface roughness \cite{vernooy_high-q_1998, ji2017ultra, Pfeiffer:18} and coupling non-ideality\cite{pfeiffer_coupling_2017, spencer2014integrated}. Impressive progress has resulted in demonstrations of high-$Q$ microresonator systems with integrated functionality \cite{liu_monolithic_2020,xiang_laser_2021}, as well as resonators that are microfabricated entirely within a CMOS foundry \cite{jin2021hertz}. With these advancements, attention has turned towards $Q$ limits imposed by the constituent photonic material themselves. For example, the presence of water, hydrogen, trace metal ions \cite{gorodetsky1996ultimate,rokhsari2004loss,liu2018ultralow,puckett2021422, pfeiffer2018ultra} and other pathways \cite{parrain2015origin,guha2017surface} are known to increase absorption. In this work, cavity-enhanced photothermal spectroscopy \cite{an_optical_1997, rokhsari2004loss,rokhsari2005observation,wang2018rapid,liu2021high,puckett2021422} is used to determine the absorption-limited $Q$ factor ($Q_{\rm abs}$) and optical nonlinearity of state-of-the-art high-$Q$ optical microresonators fabricated from four different photonic materials on silicon wafer.

Images of the microresonators characterized in this study are shown in Fig. \ref{figure1}b, where the microresonators are SiO$_2$ \cite{lee2012chemically,wu2020greater} microdisks and Si$_3$N$_4$ \cite{liu2021high}, Al$_{0.2}$Ga$_{0.8}$As \cite{chang2020ultra, xie_ultrahigh-q_2020} and Ta$_2$O$_5$ \cite{jung2021tantala} microrings. Details of the device fabrication processes are given in the Methods. Typical microresonator transmission spectra showing optical resonances are presented in Fig. \ref{figure1}b. The transmission spectra feature Lorentzian lineshapes, but in some cases are distorted by etalon effects resulting from reflection at the facets of the coupling waveguide. With such etalon effects accounted for (see Methods), the intrinsic ($Q_0$) and external (coupling) ($Q_e$) $Q$ factors can be determined. The measured intrinsic $Q_0$ factors are 418 million, 30.5 million, 2.01 million, and 2.69 million, for SiO$_2$, Si$_3$N$_4$, Al$_{0.2}$Ga$_{0.8}$As, and Ta$_2$O$_5$ devices, respectively.

The microresonator intrinsic $Q_0$ is determined by scattering and absorption losses. In order to isolate the absorption loss contribution, cavity-enhanced photothermal spectroscopy is used. The principle is based on that the resonant frequencies of dielectric microresonators are shifted by the Kerr effect and the photothermal effect, both of which result from the refractive index change that depends on the intracavity optical intensity. Because these two effects occur on very distinct time-scales (Kerr effect being ultra-fast and optical absorption occurring at a relatively slow thermal time scale from milliseconds to microseconds), it is possible to distinguish their respective contributions to resonant frequency shift and infer their nonlinear coefficients \cite{rokhsari2005observation}. Two distinct measurements are performed to determine the absorption-limited $Q_{\rm abs}$. Here, they are referred to as the ``sum measurement'' and ``ratio measurement''. In the sum measurement, resonant frequency shift is measured to obtain the sum of Kerr and photothermal effects. In the ratio measurement, the photothermal frequency response is measured to distinguish its contribution from the Kerr effect. 

\begin{figure*}
\centering
\includegraphics[width=\linewidth]{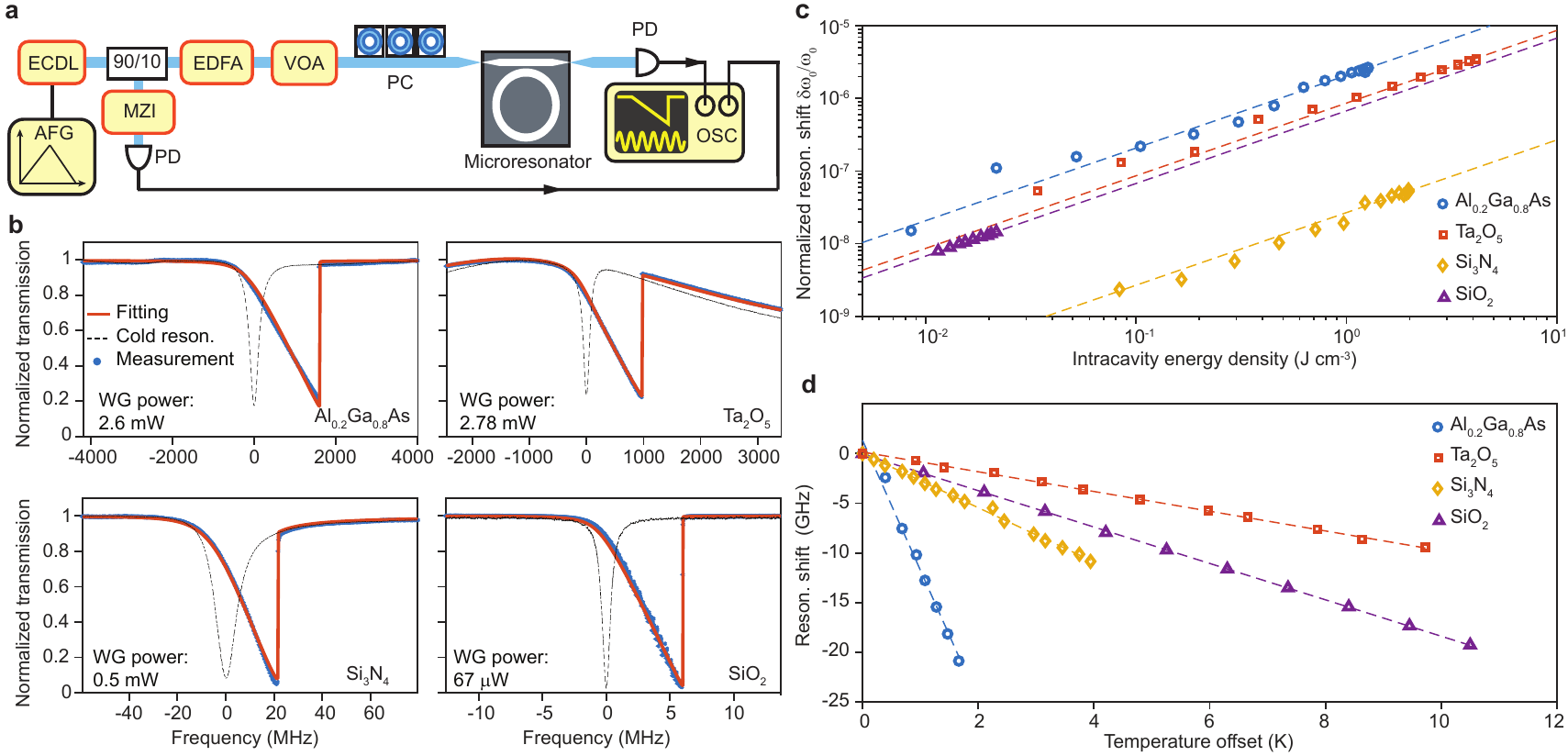}
\caption{{\bf The sum measurement.} This experiment measures the sum of Kerr and photothermal nonlinear coefficients $(g+\alpha)$. {\bf a,} Experimental setup. ECDL: external-cavity diode laser; EDFA: erbium-doped fiber amplifier; VOA: voltage-controlled optical attenuator; PC: polarization controller; PD: photodetector; MZI: Mach-Zehnder interferometer; AFG: arbitrary function generator; OSC: oscilloscope. For SiO$_2$ experiment, ECDL is replaced by a narrow-linewidth fiber laser to achieve a slower frequency tuning speed. As an aside due to narower tuning range of fiber laser, this experiment is only performed at 1550nm for SiO$_2$. {\bf b,} Typical transmission spectra of microresonators with photothermal and Kerr self-phase modulation, where the input power in the bus waveguide is indicated. Theoretical fittings are plotted in red and discussed in Methods. The cold transmission spectra measured at low pump power are also plotted with dashed lines for comparison. WG power: optical power in the bus waveguide. {\bf c,} Measured resonant frequency shift versus intracavity power for microresonators based on different materials. Dashed lines are linear fittings of the measured data. {\bf d,} Measured resonant frequency shift versus microresonator chip temperature for the four materials, with linear fittings. The fitted shift for Al$_{0.2}$Ga$_{0.8}$As, Si$_3$N$_4$, SiO$_2$ and Ta$_2$O$_5$ are -13.1, -2.84, -1.83 and -0.996, in units of GHz K$^{-1}$, respectively.
}
\label{figure2}
\end{figure*}

\begin{figure*}
\centering
\includegraphics[width=\linewidth]{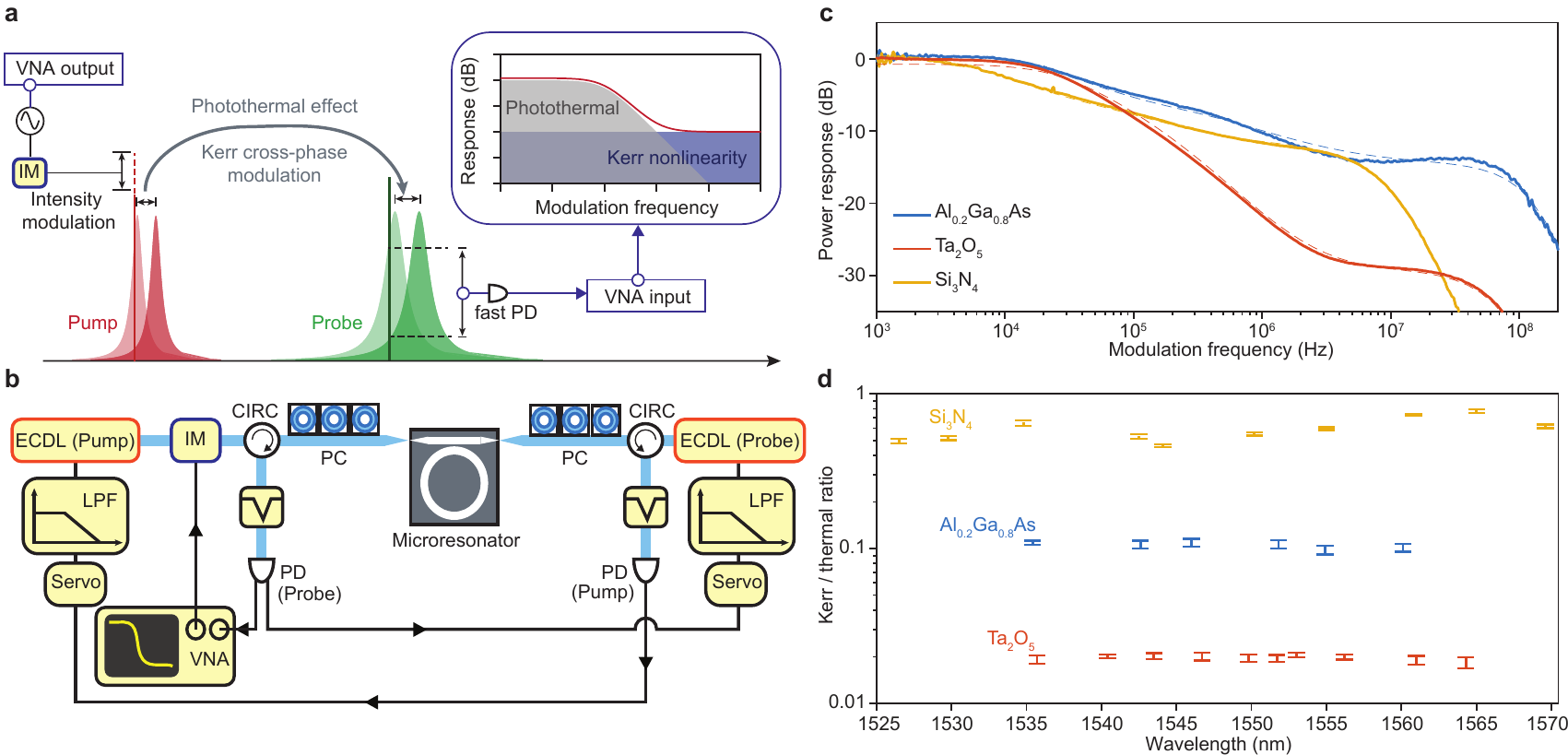}
\caption{{\bf The ratio measurement.} This experiment measures the ratio of Kerr and photothermal nonlinear coefficients $g/\alpha$. {\bf a,} Illustration of the ratio measurement. A pump laser is stabilized to a resonance and modulated by an intensity modulator. The intracavity power is thus modulated. As a result of photothermal effect and Kerr cross-phase modulation, the frequency of a nearby resonance is also modulated. Another probe laser is stabilized near this resonance, and its transmission is monitored by a vector network analyzer (VNA). Inset: the modulation response allows distinguishing the photothermal and Kerr effects. {\bf b,} Experimental setup. IM: intensity modulator; CIRC: optical circulator; LPF: low-pass filter; VNA: vector network analyzer.  {\bf c,} Typical measured response functions of the probe laser transmission $\Tilde{\mathcal{R}}$ as a function of modulation frequency $\Omega$. Numerical fittings are outlined as dashed curves. For modulation frequencies below 1 kHz, the probe response is suppressed by the servo feedback locking loop. {\bf d,} Measured wavelength dependence of the ratios between the Kerr nonlinearity and photothermal effect for three materials. 
}
\label{figure3}
\end{figure*}

In the sum measurement, the microresonator is probed by a tunable laser whose frequency is slowly swept across a resonance from the higher frequency side of a resonance (i.e., blue-detuned side). The input light polarization is aligned to the fundamental TE (Si$_3$N$_4$, Al$_{0.2}$Ga$_{0.8}$As, and Ta$_2$O$_5$) or TM (SiO$_2$) mode of the microresonator. The experimental setup is depicted in Fig. \ref{figure2}a. The frequency scan is calibrated by a radio-frequency calibrated Mach-Zehnder interferometer (MZI) \cite{li2012sideband}. The probe laser frequency scan is sufficiently slow (i.e., quasi-static scan, see Supplementary note \uppercase\expandafter{\romannumeral3} for details) to ensure that scan speed does not impact the observed lineshape through transient thermal processes within the microresonator. The transmission spectra exhibit a triangular shape \cite{carmon2004dynamical} as shown in Fig. \ref{figure2}b. Theoretical fittings of the transmission spectra are shown in red and discussed in Methods. Also, the cold resonance spectra (i.e., with very low waveguide power) measured under the same coupling conditions are plotted for comparison (dashed curve). 

By changing the input pump laser power with a voltage-controlled optical attenuator (VOA), the quasi-static resonance shift $\delta \omega_0$ of the resonant frequency $\omega_0$ versus the intracavity circulating optical energy density $\rho$ (units of J m$^{-3}$) is determined (see Supplementary note \uppercase\expandafter{\romannumeral3}) and summarized in Fig. \ref{figure2}c. The observed linear dependence contains contributions from the Kerr self-phase modulation and photothermal effects as,
\begin{equation}
    \frac{ \delta \omega_0}{\omega_0} = -\frac{1}{\omega_0}(\alpha + g) \rho,  \label{Sum}
\end{equation}
where $\alpha$ and $g$ denote the photothermal coefficient and the Kerr coefficient given by:
\begin{equation}
\begin{aligned}
     \alpha&=\overline{\kappa_a}\frac{\overline{\delta T}}{P_{\rm abs}}  \left( -\frac{\delta \omega_0}{\overline{\delta T}} \right)  V_{\rm eff},\\
     g&=\frac{ \overline{n_2} }{\overline{n_o n_g}}\omega_0 c.
\end{aligned}
\end{equation}
Here, $\kappa_a$ is the energy loss rate due to optical absorption, $n_2$ is the material Kerr nonlinear refractive index, $n_{o}$ is the material refractive index, $n_g$ is the material chromatic group refractive index, $c$ is the speed of light in vacuum, $P_{\rm abs}$ is the absorbed optical power by the microresonator and $T$ is the temperature of the microresonator. The bar (e.g., $\overline{n_2}$) denotes the average value of the underneath variable weighted by the field distribution of the optical mode. The exact definition of each average is provided in Supplementary note \uppercase\expandafter{\romannumeral1}.

The energy loss rate $\kappa_a$ is related to the material absorption-limited $Q_{\rm abs}$ factor by
\begin{equation}
    Q_{\rm abs}  = \frac{\omega_0}{\kappa_a}.
\end{equation}
To determine $\kappa_a$ and hence $Q_{\rm abs}$ from $\alpha$, it is necessary to determine $V_{\rm eff}$, ${\overline{\delta T}}/{P_{\rm abs}}$ and $\delta \omega_0/\delta T$. The effective mode volume $V_{\rm eff}$ is calculated using the optical mode obtained in finite-element modeling, and ${\overline{\delta T}}/{P_{\rm abs}}$ is further calculated using the finite-element modeling with a heat source spatially distributed as the optical mode. The resonance tuning coefficient $\delta \omega_0/\delta T$ is directly measured by varying the temperature of the microresonator chip using a thermoelectric cooler (TEC), and the results are shown in Fig. \ref{figure2}d. Since the TEC heats the entire chip, the thermo-elastic effect of the silicon substrate contributes to the frequency shift and combines with the photothermal effect. However, this thermo-elastic contribution does not appear in the sum measurement, where the heating originates only from the optical mode. Thus, the thermal-elastic contribution of the silicon substrate must be deducted from the TEC measured results (see Supplementary note \uppercase\expandafter{\romannumeral1}). Other effects that may lead to frequency shift or linewidth broadening, such as harmonic generation or multi-photon absorption, are not significant in the samples, as confirmed by observing the coupling efficiency with respect to power (see Supplementary note \uppercase\expandafter{\romannumeral3}).

\begin{figure}[t]
\centering
\includegraphics[width=\linewidth]{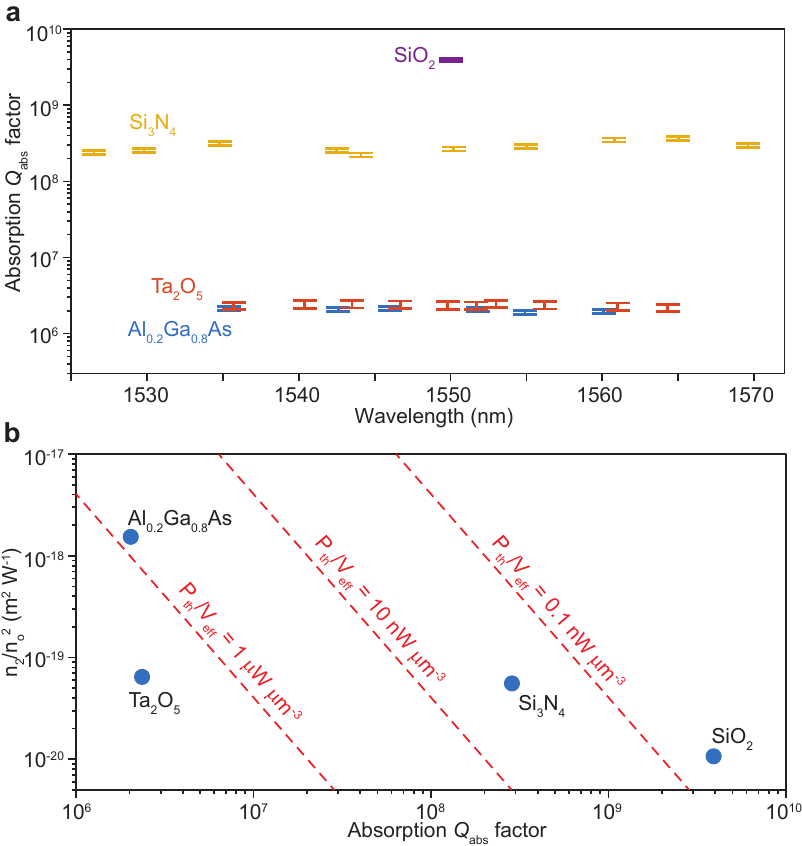}
\caption{{\bf Absorption $Q_{\rm abs}$, nonlinear coefficients and parametric oscillation threshold.} {\bf a,} Measured absorption $Q_{\rm abs}$ factors at different wavelengths in the telecommmunication C-band for the four materials. Vertical error bars give standard deviations of measurements.
{\bf b,} Comparison of absorption $Q_{\rm abs}$ factors and normalized nonlinear index ($n_2/n_o^2$) for the four materials. Measured $n_2$ values are listed in Table \ref{Table1}. The $n_2$ of SiO$_2$ was not measured here and a reported value of $2.2\times10^{-20}$ m$^2$ W$^{-1}$ is used. Parametric oscillation threshold  for a single material normalized by mode volume ($P_{\rm th}$/$V_{\rm eff}$) is indicated by the red dashed lines, assuming $\lambda = 1550$ nm, intrinsic $Q_0$ equals material absorption $Q$, and $Q_e=Q_0$ (i.e. critical coupling condition).  }
\label{figure4}
\end{figure}

\begin{table*}
\includegraphics[width=\linewidth]{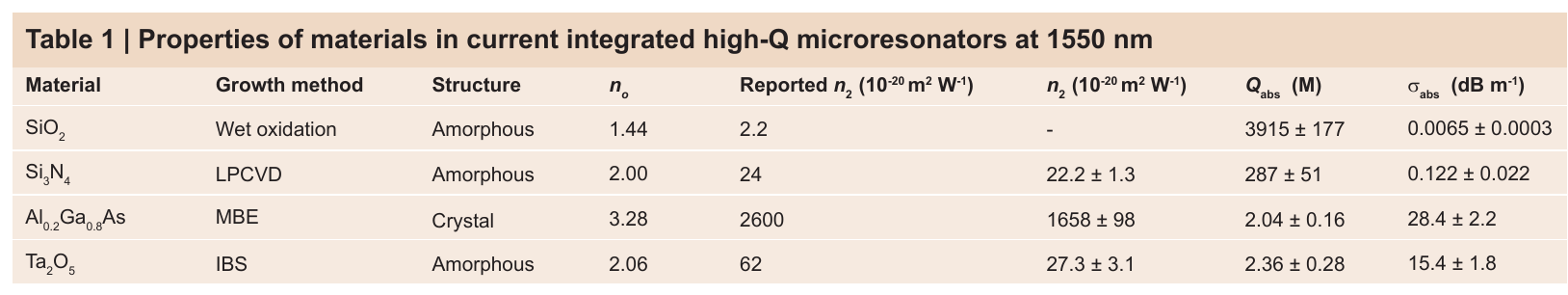}
\caption{{\bf Summary of material loss and nonlinearity.} LPCVD: low-pressure chemical vapour deposition; MBE: molecular beam epitaxy; IBS: ion-beam sputtering. Propagation loss $\sigma_{\rm abs}$ induced by absorption is calculated as $\sigma_{\rm Mat}=(10/\ln 10)\omega_0 n_g/(Q_{\rm abs} c)$.  Error indicates standard deviation. These numbers should be viewed as state-of-the-art values rather than fundamental limits.
Possible systematic errors of measurement values are discussed in Supplementary note \uppercase\expandafter{\romannumeral2}C. The $Q_{\rm abs}$ for Ta$_2$O$_5$ is further discussed in Supplementary note \uppercase\expandafter{\romannumeral3}D.
Reported $n_2$ values are taken from ref. \cite{agrawal2007nonlinear} (SiO$_2$), ref. \cite{ikeda_2008_thermal} (Si$_3$N$_4$), ref. \cite{pu_efficient_nodate} (Al$_{0.2}$Ga$_{0.8}$As) and ref. \cite{jung2021tantala} (Ta$_2$O$_5$).
 }
\label{Table1}
\end{table*}

The measurement associated with Eq. \ref{Sum} wherein the sum contributions of Kerr and photothermal effects are measured is supplemented by a measurement that provides the ratio of these quantities. This second measurement takes advantage of the very different relaxation time scales of Kerr and photothermal effects.  The experimental concept and setup are depicted in Fig. \ref{figure3}a and \ref{figure3}b. Pump and probe lasers are launched from opposite directions into the microresonator. The pump laser is stabilized close to one resonance by directly monitoring the transmission signal, and its power is modulated over a range of frequencies using a commercial lithium niobate electro-optic modulator driven by a vector network analyzer (VNA). The probe laser is locked to another nearby resonance, and is slightly detuned from the center resonant frequency. With this arrangement, pump power modulations in the first resonance induce modulations of the output probe power in the second resonance, as a result of Kerr- and photothermal-induced refractive index modulations. The probe frequency response measured for three different microresonators is presented in Fig. \ref{figure3}c. Both pump and probe laser powers are sufficiently low to minimize the thermal locking effect \cite{carmon2004dynamical}.

The plateau in the frequency response at low frequency gives the combined quasi-static contributions of photothermal and Kerr effects in the sum measurement (inset of Fig. \ref{figure3}a), while the high frequency response constitutes only the Kerr contribution. In addition, the Kerr effect here is the cross-phase modulation contribution (from the pump to the probe), while, as noted above, the Kerr self-phase modulation contribution appears in Eq. \ref{Sum}. These two effects are related by a cross-phase modulation factor $\gamma$ determined by the mode combinations used (see Methods). 
For Al$_{0.2}$Ga$_{0.8}$As and Ta$_2$O$_5$, pump and probe modes both belong to the fundamental TE mode. For Si$_3$N$_4$, pump and probe modes belong to the fundamental TE and TM modes, respectively (see Supplementary note \uppercase\expandafter{\romannumeral3}). It is noted that this measurement was challenging to perform in the suspended SiO$_2$ microdisks on account of a very slow thermal diffusion process (see Supplementary note \uppercase\expandafter{\romannumeral3}A). Instead, a published value of $n_2$ for SiO$_2$ ($2.2\times10^{-20}$ m$^2$ W$^{-1}$) was used \cite{agrawal2007nonlinear}. By numerically fitting the response curves (see Supplementary note \uppercase\expandafter{\romannumeral2}B and \uppercase\expandafter{\romannumeral3}), the ratio between Kerr and photothermal effects is extracted over a range of wavelengths and plotted in Fig. \ref{figure3}d. 

Combining results from the above sum and ratio measurements, the photothermal and Kerr coefficients are obtained individually. The inferred absorption-limited $Q_{\rm abs}$ values measured over the telecommunication C-band for each material are summarized in Fig. \ref{figure4}a. It is worth mentioning that the SiO$_2$ microdisk measurement requires a narrow-linewidth, highly-stable fiber laser on account of the microresonator's ultra-high $Q$ factor. The use of the fiber laser limits the measurement range to near 1550 nm.  A combined plot of the measured $n_2$ values (normalized by $n_o^2$) versus the absorption $Q_{\rm abs}$ is given in Fig. \ref{figure4}b (the $n_2$ of SiO$_2$ is taken from the literature \cite{agrawal2007nonlinear}). Also, in the cases of critical coupling ($Q_e=Q_0$) and absorption-limited intrinsic $Q$ factors ($Q_0=Q_{\rm abs}$), the parametric oscillation threshold per unit mode volume \cite{kippenberg2004kerr,li2012low,yi2015soliton} for a single material is shown by dashed red iso-contours:
\begin{equation}
{P_{\rm th} \over V_{\mathrm{eff}}} = {n_o^2 \, \omega_o  \over n_2 Q_{\rm abs}^2 c},
\label{Threshold}
\end{equation}
where $V_{\mathrm{eff}}$ is the effective mode volume. It should be noted that actual thresholds may be different if the optical field is not tightly confined in the core of the microresonator heterostructure.

The results described above are further summarized in Table \ref{Table1}, where, for SiO$_2$ and Si$_3$N$_4$, the measured material absorption losses are much lower than the present microresonator intrinsic losses. Therefore, improvement in microfabication of SiO$_2$ and Si$_3$N$_4$ to reduce surface roughness, hence to reduce scattering losses, will benefit photonic integrated circuits using these materials. For Al$_{0.2}$Ga$_{0.8}$As and Ta$_2$O$_5$, the material losses are close to their respective intrinsic losses, which suggests that both material and scattering loss contributions should be addressed. 

Overall, the absorption $Q_{\rm abs}$ values reported here should be viewed as state-of-the-art values that are not believed to be at fundamental limits. For example, silica glass in optical fiber exhibits loss (typically 0.2 dB km$^{-1}$) \cite{miya_ultimate_1979} that is still over one order of magnitude lower than that reported in Fig. \ref{figure4}b. 
Likewise, Ta$_2$O$_5$ is the premier material for optical coatings employed, for example, in the highest performance optical clocks and gravitational-wave interferometers. However, Ta$_2$O$_5$ exhibits fascinating stoichiometry and crystallization effects, which require careful mitigation in deposition and processing. The material-limited $Q$ of Ta$_2$O$_5$ and TiO$_2$:Ta$_2$O$_5$ has been measured to be 5 million and 25 million, respectively \cite{Pinard}. Hence, the nanofabricated devices and precision-measurement technique reported here highlight the promise to optimize material-limited performance in the Ta$_2$O$_5$ platform. It is also noted that in Al$_{0.2}$Ga$_{0.8}$As, a compound semiconductors material, surface defects may generate mid-gap states \cite{parrain2015origin} which cause extra material absorption loss. This loss mechanism will depend upon process conditions and intrinsic $Q$ factors as high as 3.52 M for Al$_{0.2}$Ga$_{0.8}$As have been reported elsewhere \cite{xie_ultrahigh-q_2020}. Finally, some of the material parameters used in modeling are impacted by factors such as the film deposition method. For example, thermal conductivity of Ta$_2$O$_5$ can depend upon the deposition method as is reflected by a wide range of values available in the literature (see Supplementary note IIID). Such effects could also impact other materials used in this study, but we have nonetheless relied upon bulk values in modeling (see Supplementary note IIC).

The current method also provides \textit{in-situ} measurement of $n_2$ for integrated photonic microresonators. We compare the $n_2$ values measured here with other reported values in Table \ref{Table1}. To give a fashion of how the nonlinearity varies between the four materials, third-order nonlinear susceptibility $\chi_{(3)}$ is calculated from the measured $n_2$ and compared with the linear susceptibility $\chi_{(1)}$. The Miller's rule \cite{miller1964optical, ettoumi2010generalized} $\chi_{(3)}\propto \chi_{(1)}^4$ relating the scaling of these two quantities is observed (see Supplementary note \uppercase\expandafter{\romannumeral4}).

In summary, the absorption loss and Kerr nonlinear coefficients of four leading integrated photonic materials have been measured using cavity-enhanced photothermal spectroscopy. The material absorption sets a practical limit of these materials in microcavity applications. The Kerr nonlinear coefficients have also been characterized, and the results are consistent with a general trend relating to nonlinearity and optical loss. Overall, the results suggest specific directions where there can be improvement in these systems as well as providing a way to predict future device performance. 

\bibliography{scibib}

\clearpage
\noindent {\bf\Large Methods} \\

\noindent{\bf Fabrication of optical microresonators.} 
The SiO$_2$ microresonator is fabricated by thermally growing 8-$\mu$m thick thermal wet oxide on a 4 inch float-zone silicon wafer, followed by i-line stepper photolithography, buffered oxide etch, XeF$_2$ silicon isotropic dry etch and thermal annealing \cite{lee2012chemically,wu2020greater}. The Si$_3$N$_4$ microresonator is fabricated with the photonic Damascene process, including using deep-ultraviolet stepper lithography, preform etching, low-pressure chemical vapour deposition, planarization, cladding and annealing \cite{liu2021high}. The Al$_{0.2}$Ga$_{0.8}$As microresonator is fabricated with an epitaxial Al$_{0.2}$Ga$_{0.8}$As layer bonded onto a silicon wafer with a 3-$\mu$m thermal SiO$_2$ layer, followed by GaAs substrate removal, deep ultraviolet patterning, inductively coupled plasma etching, passivation with Al$_2$O$_3$ and SiO$_2$ cladding \cite{chang2020ultra, xie_ultrahigh-q_2020}. The Ta$_2$O$_5$ microresonator is fabricated by ion-beam sputtering Ta$_2$O$_5$ deposition followed by annealing, electron-beam lithography, Ta$_2$O$_5$ etching, ultraviolet lithography and dicing \cite{jung2021tantala}. 

\medskip

\noindent{\bf Experimental details.} 
In the sum measurement, the scanning speed of the laser frequency is decreased until the mode's broadening as induced by the thermo-optic shift becomes stable (i.e., not influenced by the scan rate). Also, the waveguide input power is minimized such that it is well below the threshold of parametric oscillation. The power is calibrated using the photodetector voltage.

In the ratio measurement, the optical frequencies of the pump and probe lasers are locked to their respective cavity modes using a servo feedback with 1 kHz bandwidth. The pump laser is locked near the mode resonant frequency, while the probe laser is locked to the side of the resonance to increase transduction of refractive index modulation into transmitted probe power. The intensity modulator is calibrated in a separate measurement under the same driving power.

\medskip

\noindent{\bf Fitting of spectra in the sum measurement.} 
For Si$_3$N$_4$ and Ta$_2$O$_5$ devices, the transmission spectrum is the interference of a Lorentzian-lineshaped mode resonance with a background field contributed by facet reflections of the waveguide. The transmission function of a cavity resonance is given by 
\begin{equation}
    T_{\rm res}
    =1-\frac{\kappa_e}{\kappa/2+i[\Delta-(\alpha+g)\rho ]}, 
    \label{Eq: Tr}
\end{equation}
where $\Delta$ is the cold-cavity laser-cavity detuning, $\alpha$ and $g$ are the absorption and Kerr nonlinear coefficients, respectively, and $\rho $ is the intracavity energy density as defined in the main text. The reflection at the two waveguide facets forms a low-finesse Fabry–P\'{e}rot resonator. Combining this waveguide reflection with the cavity resonance, the overall amplitude transmission is given by (see Supplementary note \uppercase\expandafter{\romannumeral2})
\begin{equation}
    T\propto
    \left|\frac{T_{\rm res}}{1-r T_{\rm res}^2 \exp[i(-\Delta/\omega_{\rm FP}+\phi)]}\right|^2, 
    \label{transmission}
\end{equation}
where $r$ is the reflectivity at the waveguide facet, $\omega_{\rm FP}$ is the free spectral range of the facet-induced Fabry–P\'{e}rot cavity (in rad/s units), and $\phi$ is a constant phase offset. 

In the experiment, the above quantities are fitted in three steps. First, $\omega_{\rm FP}$ and $r$ are obtained by measuring the transmission away from mode resonances. Next, loss rates $\kappa$ and $\kappa_{e}$ can be determined by measuring the transmission of the mode at a low probe power. Finally, launching higher power into the microresonator allows the mode broadening to be observed and the transmission is fitted with Eq. (\ref{transmission}), where $(\alpha+g)$ is the fitting variable and other parameters are obtained from the previous steps. For Al$_{0.2}$Ga$_{0.8}$As and SiO$_2$ devices which have no Fabry–P\'{e}rot background, $r$ can be set to zero and the first step in the above fitting procedure can be omitted. The fitting results are presented in Fig. \ref{figure2}b.

\medskip

\noindent{\bf Fitting of response in the ratio measurement.}
The response of the probe mode resonant frequency $\Tilde{\delta}_{\rm b}$ as a result of pump power modulation $\Tilde{P}_{ \mathrm{in}}$ can be described by (see Supplementary note \uppercase\expandafter{\romannumeral2}),
\begin{equation}
    \frac{\Tilde{\delta}_{\rm b}(\Omega)}{  \Tilde{P}_{ \mathrm{in}  } (\Omega) } = -\frac{\alpha \Tilde{r}(\Omega)+\gamma g}{V_{\rm eff}}\frac{2\eta_{\rm p}}{i\Omega+\kappa_{\rm p}/2}.
    \label{Eq: first order resonance shift, probe mode}
\end{equation}
where $\Omega$ is the pump power modulation frequency (in rad/s units), $\Tilde{P}_{\mathrm{in}}$ is the modulation amplitude of the pump power, $\kappa_{\rm p}$ is the total loss rate of the pump mode, $\eta_{\rm p} = \kappa_{e,\rm p}/\kappa_{\rm p}$ is the coupling efficiency for pump mode, $\alpha$ is the absorption coefficient as mentioned in the previous section, $\Tilde{r}$ is the frequency response of modal temperature modulation as a result of thermal diffusion, and the factor $\gamma$ accounts for cross-phase modulation of the probe mode by the pump mode.

The frequency response of the transmitted probe mode with respect to its resonance shift $\Tilde{\delta}_{\rm b}(\Omega)$ is derived in Supplementary note \uppercase\expandafter{\romannumeral2} and has the following form:
\begin{equation}
    \frac{\Tilde{T}_{\rm b}(\Omega)}{\Tilde{\delta}_{\rm b}(\Omega)} = -\frac{2 \kappa_{e,\rm b} \Delta_{\rm b}^{(0)}}{\kappa_{\rm b}^2/4+\left(\Delta_{\rm b}^{(0)}\right)^2}\frac{\kappa_{\rm b}-\kappa_{e,\rm b}+i\Omega}{(\kappa_{\rm b}/2+i\Omega)^2+\left(\Delta_{\rm b}^{(0)}\right)^2}|a_{\mathrm{in, b}}|^2,
    \label{Eq: response formula final form}
\end{equation}
where $\Delta_{\rm b}^{(0)}$ is the steady-state detuning of the probe mode when no modulation is present, and $\kappa_{\rm b}$ and $\kappa_{e,\rm b}$ refer to the total loss rate and external coupling rate for the probe mode.

The response curve in Fig. \ref{figure3}c is modeled by,
\begin{equation}
    \Tilde{\mathcal{R}}(\Omega)=
    \frac{\Tilde{T}_{\rm b}(\Omega)}{\Tilde{P}_{\rm in}(\Omega)}=
    \frac{\Tilde{T}_{\rm b}(\Omega)}{\Tilde{\delta}_{\rm b}(\Omega)}
    \frac{\Tilde{\delta}_{\rm b}(\Omega)}{\Tilde{P}_{\rm in}(\Omega)}. 
\end{equation}
and is fitted according to Eq. \eqref{Eq: first order resonance shift, probe mode} and \eqref{Eq: response formula final form}. In the fitting, $\kappa$ and $\kappa_e$ have been measured separately, $\Tilde{r}$ is determined from finite element method simulations, and the probe mode $\Delta_0$ and ratio $\alpha/g$ are parameters to be fitted.

\medskip
\noindent {\bf\large Data availability} \\
\noindent All data generated or analysed during this study are available within the paper and its supplementary materials. Further source data will be made available upon reasonable request.\\

\noindent {\bf\large Code availability} \\
\noindent The analysis codes will be made available upon reasonable request.\\

\noindent {\bf Acknowledgments} The authors thank H. Blauvelt and Z. Yuan for helpful comments, Z. Liu for discussions on data processing, as well as D. Carlson for preparing the Ta$_2$O$_5$ SEM image. Q.-X. J. acknowledges the Caltech Student Faculty Program for financial supports. This work was supported by DARPA under the DODOS (HR0011-15-C-0055) and the APHi (FA9453-1 9-C-0029) programs.\\
\\
\noindent{\bf Author contributions} \\
\noindent M.G., Q.-F.Y., Q.-X.J. and J.L. conceived the idea. 
M.G., Q.-F.Y. and Q.-X.J. performed the measurement with assistance from H.W., L.W., and B.S.. 
M.G., Q.-F.Y., Q.-X.J. and H.W. devised the theoretical model. 
L.W. provided the SiO$_2$ samples. 
J.L. and G.H. provided the Si$_3$N$_4$ samples. 
L.C. and W.X. provided the Al$_{0.2}$Ga$_{0.8}$As samples. 
S.-P.Y. provided the Ta$_2$O$_5$ samples. 
All the authors analyzed the data and wrote the manuscript. K.J.V., T.J.K., J.E.B. and S.B.P. supervised the project. 
\\
\\
\noindent{\bf Competing interests} The authors declare no competing financial interests.\\
\\
\noindent{\bf Additional information} \\
\noindent{\bf Supplementary information} is available for this paper.\\
\noindent{\bf Correspondence and requests for materials} should be addressed to K.J.V.

\end{document}